\documentclass[twocolumn,prb,showpacs]{revtex4}
\usepackage{psfrag}
\usepackage{graphicx}
\usepackage{amsmath}
\usepackage{amsmath, amssymb, bm}
\usepackage{eufrak}

\newcommand{\br}{\bm{r}}
\newcommand{\bra}[1]{\langle #1 |}
\newcommand{\ket}[1]{| #1 \rangle}
\newcommand{\braket}[2]{\langle #1 | #2 \rangle}
\newcommand{\dr}{\mathrm{d}\bm{r}}
\newcommand{\rr}{|\bm{r} - \bm{r}'|}

\DeclareMathOperator{\Tr}{Tr}

\begin{document}
\title{Fully selfconsistent GW calculations for molecules}

\author{C. Rostgaard, K. W. Jacobsen, and K. S. Thygesen}
\affiliation{Center for Atomic-scale Materials Design (CAMD), \\
  Department of Physics, Technical University of Denmark, DK - 2800
  Kgs. Lyngby, Denmark}
\date{\today}

\begin{abstract}
  We calculate single-particle excitation energies for a series of 33
  molecules using fully selfconsistent GW, one-shot G$_0$W$_0$,
  Hartree-Fock (HF), and hybrid density functional theory (DFT). All
  calculations are performed within the projector augmented wave (PAW)
  method using a basis set of Wannier functions augmented by numerical
  atomic orbitals. The GW self-energy is calculated on the real
  frequency axis including its full frequency dependence and
  off-diagonal matrix elements.  The mean absolute error of the
  ionization potential (IP) with respect to experiment is found to be
  4.4, 2.6, 0.8, 0.4, and 0.5 eV for DFT-PBE, DFT-PBE0, HF, G$_0$W$_0$[HF], and selfconsistent GW, respectively.  This
  shows that although electronic screening is weak in molecular
  systems its inclusion at the GW level reduces the error in the IP by
  up to 50\% relative to unscreened HF. In general GW overscreens
  the HF energies leading to underestimation of the IPs. The best IPs
  are obtained from one-shot G$_0$W$_0$ calculations based on HF since this reduces the overscreening. Finally, we find that the inclusion of
  core-valence exchange is important and can affect the
  excitation energies by as much as 1 eV.
\end{abstract}

\maketitle 

\section{Introduction}
Density functional theory (DFT)\cite{hohenbergkohn} with the
single-particle Kohn-Sham (KS) scheme\cite{kohnsham} is today the most
widely used approach to the electronic structure problem of real
materials in both solid state physics and quantum chemistry.  While
properties derived from total energies (or rather total energy
differences) are accurately predicted by DFT, it is well known that
DFT suffers from a band gap problem implying that the single-particle
KS eigenvalues cannot in general be interpreted as real quasiparticle
(QP) excitation energies. In particular, semilocal
exchange-correlation functionals severely underestimate the
fundamental gap of both insulators, semi-conductors, and
molecules.\cite{hybertsen86,perdew_zunger,paier06,heyd05}

The hybrid\cite{becke,LYP,pbe0} and screened hybrid\cite{hse03}
functionals, which admix around 25$\%$ of the (screened) Fock exchange
with the local DFT exchange, generally improve the description of band
gaps in bulk semi-conductors and insulators\cite{paier06,heyd05}.
However, the orbital energies obtained for finite systems using such
functionals still underestimate the fundamental gap, $I_p-E_a$, (the
difference between ionization potential and electron affinity) by up
to several electron volts. In fact, for molecules the pure
Hartree-Fock (HF) eigenvalues are usually closer to the true electron
addition/removal energies than are the hybrid DFT eigenvalues. This is
because HF is self-interaction free and because screening of the
exchange interaction is a relatively weak effect in molecular
systems.\cite{perdew_zunger,pemmaraju07,jarnak_comment}.  On the other
hand, in extended systems the effect of self-interaction is less
important and the long range Coulomb interaction becomes short ranged
due to dynamical screening.  As a consequence HF breaks down in
extended systems leading to dramatically overestimated band gaps and a
qualitatively incorret description of
metals.\cite{hf_band_polyacethylene,grosso,hf_gap_Ar}

The many-body GW approximation of Hedin\cite{hedin} has been widely
and succesfully used to calculate QP band structures in metals,
semi-conductors, and
insulators.\cite{hybertsen86,gunnarsson,wilkins,orr} The GW
approximation can be viewed as HF with a dynamically screened Coulomb
interaction. The fact that the screening is determined by the system
itself instead of being fixed a priori as in the screened hybrid
schemes, suggests that the GW method should be applicable to a broad
class of systems ranging from metals with strong screening to
molecules with weak screening.  With the entry of nanoscience the use
of GW has been extended to low-dimensional systems and
nanostructures\cite{stan_eulett,grossman01,hahn,tb_dft,louie,reining,Chelikowsky,neaton,thygesen09,Freysoldt,juanma_prb}
and more recently even nonequilibrium phenomena like quantum
transport\cite{thygesen_jcp,thygesen_prl,leeuwen_trans,spataru2,verdozzi}.
In view of this trend it is important to establish the performance of
the GW approximation for other systems than the crystalline solids. In
this work we present first-principles benchmark GW calculations for a
series of small molecules. In a closely related study we compared GW
and Hartree-Fock to exact diagonalization results for semi-empirical
PPP models of conjugated molecules\cite{paperI}. The main conclusions
from the two studies regarding the qualities of the GW approximation
in molecular systems, are very consistent.

Most GW calculations to date rely on one or several approximations of
more technical character. These include the plasmon pole
approximation, the linearized QP equation, neglect of off-diagonal
matrix elements in the GW self-energy, analytic continuations from the
imaginary to the real frequency axis, neglect of core states
contributions to the self-energy, neglect of self-consistency. The
range of validity of these approximations has been explored for solid
state systems by a number of authors\cite{barth,usuda_hamada,ku,schilfgaarde_prb,Shishkin2006,rinke}, however, much less is known about their applicability to molecular systems\cite{stan_eulett}. Our
implementation of the GW method avoids all of these technical
approximations allowing for a direct and unbiased assessment of the GW
approximation itself.

Here we report on single-shot G$_0$W$_0$ and fully self-consistent GW
calculations of QP energies for a set of 33 molecules.  The calculated
IPs are compared with experimental values as well as single-particle
eigenvalues obtained from Hartree-Fock and DFT-PBE/PBE0 theories. As
additional benchmarks we compare to second-order M{\"o}ller-Plesset
(MP2) and DFT-PBE total energy differences between the neutral and
cation species.  Special attention is paid to the effect of
selfconsistency in the GW self-energy and the role of the initial Green function,
$G_0$, used in one-shot G$_0$W$_0$ calculations. The use of PAW
rather than pseudopotentials facilitate the inclusion of core-valence
exchange, which we find can contribute significantly to the HF and GW energies.
Our results show that the GW approximation yields accurate
single-particle excitation energies for small molecules improving both
hybrid DFT and full Hartree-Fock results. 

The paper is organized as follows. In Sec. \ref{sec.method} we
describe the theoretical and numerical details behind the GW
calculations, including the augmented Wannier function basis set, the
self-consistent solution of the Dyson equation, and the evaluation of
valence-core exchange within PAW. In Sec. \ref{sec.results} we discuss
and compare the results of G$_0$W$_0$, GW, HF, PBE0, and PBE
calculations. We analyze the role of dynamical screening, and discuss
the effect of self-consistency in the GW self-energy. We conclude in
Sec. \ref{sec.conclusions}.

\section{Method}\label{sec.method}
\subsection{Augmented Wannier function basis}
For the GW calculations we apply a basis set consisting of projected
Wannier functions (PWF) augmented by numerical atomic orbitals (NAO).
The PWFs, $\phi_i$, are obtained by maximizing their projections onto
a set of target NAOs, $\Phi_{Alm}$, subject to the condition that they
span the set of occupied eigenstates, $\psi_n$. Thus we maximize the
functional
\begin{equation}
\Omega = \sum_{i}\sum_{A,l,m}|\langle \phi_i|\Phi_{Alm}\rangle|^2
\end{equation}
subject to the condition $\text{span}\{\phi_i\}\supseteq
\text{span}\{\psi_n\}_{\text{occ}}$ as described in
Ref.~\onlinecite{quambo}. The target NAOs are given by
$\Phi_{Alm}(\bold r)=\zeta_{Al}(r)Y_{lm}(\bold r)$ where $\zeta_{Al}$
is a modified Gaussian which vanish outside a specified cut-off
radius, and $Y_{lm}$ are the spherical harmonics corresponding to the
valence of atom $A$. The number of PWFs equals the number of target
NAOs. For example we obtain one PWF for H ($l_{\text{max}}=0$), and
four PWFs for C ($l_{\text{max}}=1$).  The PWFs mimick the target
atomic orbitals but in addition they allow for an \emph{exact}
representation of all the occupied molecular eigenstates. The latter
are obtained from an accurate real-space PAW-PBE
calculation\cite{GPAW,GPAW-LCAO}.

The PWFs obtained in this way provide an exact representation of the
occupied PBE eigenstates. However, this does not suffice for GW
calculations because the polarizability, $P$, and the screened
interaction, $W$, do not live in this subspace.  Hence we augment the
PWFs by additional NAOs including so-called polarization functions
which have $l=l_{\text{max}}+1$ and/or extra radial functions (zeta
functions) for the valence atomic orbitals. For more details on the definition of
polarization- and higher zeta functions we refer to Ref. \onlinecite{GPAW-LCAO}. To give an example, a double-zeta-polarized (DZP)
basis consists of the PWFs augmented by one set of NAOs corresponding
to $l=0,...,l_{\text{max}}$ and one set of polarization orbitals.
Note that the notation, SZ, SZP, DZ, DZP, etc., is normally used for
pure NAO basis sets, but here we use it to denote our augmented
Wannier basis set.  We find that the augmented Wannier basis is
significantly better for HF and GW calculations than the corresponding
pure NAO basis.

The GW and HF calculations presented in Sec. \ref{sec.results} were
performed using a DZP augmented Wannier basis. This gives a total of 5
basis functions per H, Li, and Na, and 13 basis functions for all
other chemical elements considered. In Sec. \ref{sec.conv} we discuss
convergence of the GW calculations with respect to the size of the
augmented Wannier basis.

\subsection{GW calculations}
The HF and GW calculations for isolated molecules are performed using
a Green function code developed for quantum
transport.\cite{thygesen_gw_prb} In principle, this scheme is designed
for a molecule connected to two electrodes with different chemical
potentials $\mu_L$ and $\mu_R$.  However, the case of an isolated
molecule can be treated as a special case by setting $\mu_L=\mu_R=\mu$
and modelling the coupling to electrodes by a constant imaginary
self-energy, $\Sigma_{L/R}=i\eta$. The chemical potential $\mu$ is
chosen to lie in the HOMO-LUMO gap of the molecule and the size of
$\eta$, which provides an artifical broadening of the discrete levels,
is reduced until the results have converged. In this limit of small
$\eta$ the result of the GW calculation becomes independent of the
precise position of $\mu$ inside the gap.

In Ref.~\onlinecite{thygesen_gw_prb} the GW-transport scheme was
described for the case of an orthogonal basis set and for a truncated,
two-index Coulomb interaction. Below we generalize the relevant
equations to the case of a non-orthogonal basis and a full four-index
Coulomb interaction. Some central results of many-body perturbation
theory in a non-orthogonal basis can be found in Ref.
\onlinecite{nonorthogonal}.
  
The central object is the retarded Green function, $G^r$,
\begin{equation}\label{eq.gr}
  G^r(\varepsilon)=[(\varepsilon+i\eta)S-H_{\text{KS}}+v_{\text{xc}}-\Delta
v_{\text{H}}-\Sigma_{\text{xc}}^r[G](\varepsilon)]^{-1}
\end{equation}
In this equation all quantities are matrices in the augmented Wannier
basis, e.g.  $H_{\text{KS},ij}=\langle \phi_i|\hat
H_{\text{KS}}|\phi_j\rangle$ is the KS Hamiltonian matrix and
$S_{ij}=\langle \phi_i|\phi_j\rangle$ is an overlap matrix. The term
$\Delta v_{\text{H}}$ represents the change in the Hartree potential
relative to the DFT Hartree potential already contained in
$H_{\text{KS}}$, see Appendix \ref{sec:GWselfenergy}.  The local
xc-potential, $v_{\text{xc}}$, is subtracted to avoid double counting
when adding the many-body self-energy, $\Sigma_{\text{xc}}[G]$. As
indicated, the latter depends on the Green function and therefore Eq.
(\ref{eq.gr}) must in principle be solved self-consistently in
conjuction with the self-energy.

In the present study $\Sigma_{\text{xc}}$ is either the bare exchange
potential or the GW self-energy. To be consistent with the code used
for the calculations, we present the equations for the GW self-energy
on the so-called Keldysh contour.  However, under the equilibrium
conditions considered here the Keldysh formalism is equivalent to the
ordinary time-ordered formalism.

The GW self-energy is defined by
\begin{equation}
  \label{eq:GWKeldysh}
  \Sigma^{\text{GW}}_{ij}(\tau, \tau') = i\sum_{kl}G_{kl}(\tau, \tau'^+)W_{ik,jl}
(\tau,\tau'),
\end{equation}
where $\tau$ and $\tau'$ are times on the Keldysh contour, $\mathcal
C$.  The dynamically screened Coulomb interaction obeys the Dyson-like
equation
\begin{multline}
  W_{ij,kl}(\tau,\tau') = V_{ij,kl} \delta_\mathcal{C}(\tau,\tau')\\ + \sum_{pqrs}
\int_{\mathcal C}\! \mathrm{d}\tau_1 V_{ij,pq}P_{pq,rs}(\tau,\tau_1)
W_{rs,kl}(\tau_1,\tau'),
\end{multline}
and the polarization bubble is given by
\begin{equation}
  P_{ij,kl}(\tau,\tau') = -iG_{ik}(\tau,\tau')G_{lj}(\tau',\tau).
\end{equation}
In the limit of vanishing polarization, $P=0$, $W$ reduces to the bare
Coulomb interaction
\begin{equation} \label{eq:coulomb-elements}
  V_{ij,kl} = \iint\! \frac{\dr\,\dr'}{\rr}
\phi_i(\bm{r}) \phi^*_j(\bm{r})\phi_k^*(\bm{r}') \phi_l(\bm{r}')
\end{equation}
and the GW self-energy reduces to the exchange potential of HF theory.

From the above equations for the contour-ordered quantities, the
corresponding real time components, i.e. the retarded, advanced,
lesser, and greater components, can be obtained from standard
conversion rules\cite{haug_jauho,leeuwen_keldysh}. For completeness we
give the expressions for the real time components of the GW equations
in Appendix \ref{sec:GWselfenergy}.

The time/energy dependence of the dynamical quantities $G$, $W$, $P$,
and $\Sigma$, is represented on a uniform grid. We switch between time
and energy domains using the Fast Fourier Transform in order to avoid
time consuming convolutions. A typical energy grid used for the GW
calculations in this work ranges from -150 to 150 eV with a grid
spacing of 0.02 eV. The code is parallelized over basis functions and
energy grid points.  We use a Pulay mixing scheme for updating the
Green function $G^r$ when iterating Eq. \eqref{eq.gr} to
self-consistency as described in Ref. \onlinecite{thygesen_gw_prb}.

We stress that no approximation apart from the finite basis set is
made in our implementation of the GW approximation. In particular the
frequency dependence is treated exactly and analytic continuations
from the imaginary axis are avoided since we work directly on the real
frequency/time axis. The price we pay for this is the large size of
the energy grid.

\subsection{Spectral function}
The single-particle excitation spectrum is contained in the spectral function
\begin{equation}
  \label{eq:spectral function}
  A(\varepsilon) = i (G^r(\varepsilon) - [G^r(\varepsilon)]^\dagger).
\end{equation}
For a molecule $A(\varepsilon)$ shows peaks at the QP energies
$\varepsilon_n=E_n(N+1)-E_0(N)$ and $\varepsilon_n=E_0(N)-E_n(N-1)$
corresponding to electron addition and removal energies, respectively.
Here $E_n(N)$ denotes the energy of the $n$th excited state of the
system with $N$ electrons and $N$ refers to the neutral state.
 
When the Green function is evaluated in a non-orthogonal basis, like
the augmented Wannier basis used here, the projected density of states
for orbital $\phi_i$ becomes
\begin{equation}
  D_i(\varepsilon) = [S A(\varepsilon) S]_{ii} ~ / ~ 2\pi
S_{ii},
\end{equation}
where matrix multiplication is implied.\cite{nonorthogonal}
Correspondingly, the total density of states, or quasiparticle
spectrum, is given by
\begin{equation}\label{eq.dos}
  D(\varepsilon) = \Tr (A(\varepsilon) S) / 2 \pi.
\end{equation}

\subsection{Calculating Coulomb matrix elements}
The calculation of all of the Coulomb matrix elements, $V_{ij,kl}$, is
prohibitively costly for larger basis sets.  Fortunately the matrix is
to a large degree dominated by negligible elements. To systematically
define the most significant Coulomb elements, we use the product basis
technique of Aryasetiawan and Gunnarsson
\cite{aryasetiawan,stan_levels}. In this approach, the pair orbital
overlap matrix
\begin{equation}
  \label{eq:pairorb-overlap}
  S_{ij,kl} = \braket{n_{ij}}{n_{kl}},
\end{equation}
where $n_{ij}(\bold r)=\phi_i^*(\bold r)\phi_j(\bold r)$
is used to screen for the significant elements of $V$.

The eigenvectors of the overlap matrix Eq. \eqref{eq:pairorb-overlap}
represents a set of ``optimized pair orbitals'' and the eigenvalues
their norm. Optimized pair orbitals with insignificant norm must also
yield a reduced contribution to the Coulomb matrix, and are omitted in
the calculation of $V$. We limit the basis for $V$ to optimized pair
orbitals with a norm larger than $10^{-5} a_0^{-3}$. This gives a
significant reduction in the number of Coulomb elements that needs to
be evaluated, and it reduces the matrix size of $P(\varepsilon)$ and
$W(\varepsilon)$ correspondingly, see Appendix \ref{sec:GWselfenergy}.

The evaluation of the double integral in Eq.
\eqref{eq:coulomb-elements} is efficiently performed in real space by
solving a Poisson equation using multigrid
techniques\cite{GPAW,Walter2008}.

\subsection{Valence-core exchange}
All inputs to the GW/HF calculations, i.e.  the selfconsistent
Kohn-Sham Hamiltonian, $H_{\text{KS}}$, the xc potential
$v_\text{xc}$, the Coulomb matrix elements, $V_{ij,kl}$, are
calculated using the real-space PAW\cite{bloechl} code
GPAW\cite{GPAW,GPAW-LCAO}.

In GPAW, the core electrons (which are treated
scalar-relativistically) are frozen into the orbitals of the free
atoms, and the Kohn-Sham equations are solved for the valence states
only. Unlike pseudo potential schemes, these valence states are
subject to the full potential of the nuclei and core electrons. This
is achieved by a partitioning scheme, where quantities are divided
into pseudo components augmented by atomic corrections. The operators
obtained from GPAW are thus full-potential quantities, and the wave
functions from which the Wannier basis functions are constructed
correspond to the all-electron valence states. Ref.
\onlinecite{Walter2008} describes how the all-electron Coulomb
elements can be determined within the PAW formalism.

Since both core and all-electron valence states are available in the
PAW method, we can evaluate the contribution to the valence exchange
self-energy coming from the core electrons. As the density matrix is
simply the identity matrix in the subspace of atomic core states, this
valence-core exchange reads
\begin{equation}
  \label{eq:valence-core}
\Sigma_{x,ij}^{\text{core}}=  -\sum_k^\text{core} V_{ik,jk},
\end{equation}
where $i, j$ represent valence basis functions. We limit the inclusion
of valence-core interactions to the exchange potential, neglecting it
in the correlation. This is reasonable, because the polarization
bubble, $P$, involving core and valence states will be small due to
the large energy difference and small spatial overlap of the valence
and core states.  This procedure was used and validated for solids in
Ref.  \onlinecite{Shishkin2006}. We find that the elements of
$\Sigma_{x,ij}^{\text{core}}$ can be significant -- on average 1.2 eV
for the HOMO -- and are larger (more negative) for the more bound
orbitals which have larger overlap with the core states. In general,
the effect on the HOMO-LUMO gap is to enlarge it, on average by 0.4 eV
because the more bound HOMO level is pushed further down than the less
bound LUMO state. In the case of solids, the role of valence-core
interaction has been investigated by a number of
authors\cite{usuda_hamada, ku, schilfgaarde_prb, Shishkin2006, engel}.
Here the effect on the QP band gap seems to be smaller than what we
find for the molecular gaps.  We note that most GW calculations rely
on pseudopotential schemes where these valence-core interactions are
not accessible. In such codes, the xc contribution from the core
electrons are sometimes estimated by $\Sigma^\text{core}_\text{xc}
\approx v_\text{xc}[n] - v_\text{xc}[n_\text{val}]$ where
$n_\text{val}$ is the valence electron density, but as the local xc
potential is a non-linear functional of the density, this procedure is
not well justified. Instead we subtract the xc potential of the full
electon density $n$, and add explicitly the exact exchange core
contribution.

\section{Results}\label{sec.results}
\begin{figure}[!ht]
\includegraphics[width=0.9\linewidth]{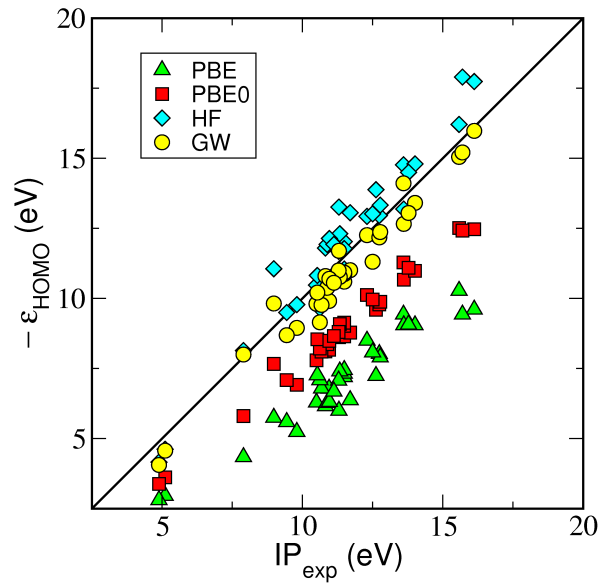}
\caption[cap.wavefct]{\label{fig.ip} (Color online) Calculated
  negative HOMO energy versus experimental ionization potential. Both
  PBE and PBE0 systematically understimates the ionization energy due
  to self-interaction errors while HF overestimates it slightly. The
  dynamical screening from the GW correlation lowers the HF energies
  bringing them closer to the experimental values. Numerical values
  are listed in Table \ref{tab.num}.}
\end{figure}

In Fig. \ref{fig.ip} we compare the calculated HOMO energies with
experimental ionization potentials for the 33 molecules listed in
Table \ref{tab.num}. The geometries of the molecules, which all belong
to the G2 test set, are taken from Ref.~\onlinecite{geom}. The
different HOMO energies correspond to: DFT-PBE\cite{pbe} and
DFT-PBE0\cite{pbe0} eigenvalues, Hartree-Fock eigenvalues, and 
fully selfconsistent GW. The GW energies are obtained from the peaks in the corresponding density of
states Eq. (\ref{eq.dos}) extrapolated to $\eta=0$ ($\eta$ gives an
artificial broadening of the delta peaks). 

We stress the different meaning of fully selfconsistent GW and the
recently introduced method of quasiparticle selfconsistent
GW\cite{schilfgaarde_qsGW}: In fully selfconsistent GW the Green
function obtained from Dyson's equation Eq. (\ref{eq.gr}) with
$\Sigma_{xc}[G]=\Sigma_{\text{GW}}[G]$ is used to calculate the
$\Sigma_{\text{GW}}$ of the next iteration. In QP-selfconsistent GW,
$\Sigma_{\text{GW}}$ is always evaluated using a non-interacting Green
function and the self-consistency is obtained when the difference
between the non-interacting GF and the interacting GF, is minimal.

Fig. \ref{fig.ip} clearly shows that both the PBE and PBE0 eigenvalues
of the HOMO severely underestimates the ionization potential. The
average deviation from the experimental values are 4.35 eV and 2.55
eV, respectively. The overestimation of the single-particle
eigenvalues of occupied states is a well known problem of DFT and can
be ascribed to the insufficient cancellation of the self-interaction
in the Hartree potential.\cite{perdew_zunger,pemmaraju07} Part of this
self-interaction is removed in PBE0. However, the fact that the HF
results are significantly closer to experiments indicates that the
25\% Fock exchange included in the PBE0 is not sufficient to cure the
erroneous description of (occupied) molecular orbitals.  On the other
hand PBE0 gives good results for band gaps in semi-conductors and
insulators where in contrast full Hartree-Fock does not perform
well.\cite{hf_band_polyacethylene,grosso,hf_gap_Ar} We conclude that
the amount of Fock exchange to be used in the hybrid functionals to
achieve good quasiparticle energies is highly system dependent. A
similar problem is encountered with self-interaction corrected
exchange-correlation functionals.\cite{pemmaraju07}

\begin{figure}[!ht]
\includegraphics[width=0.9\linewidth]{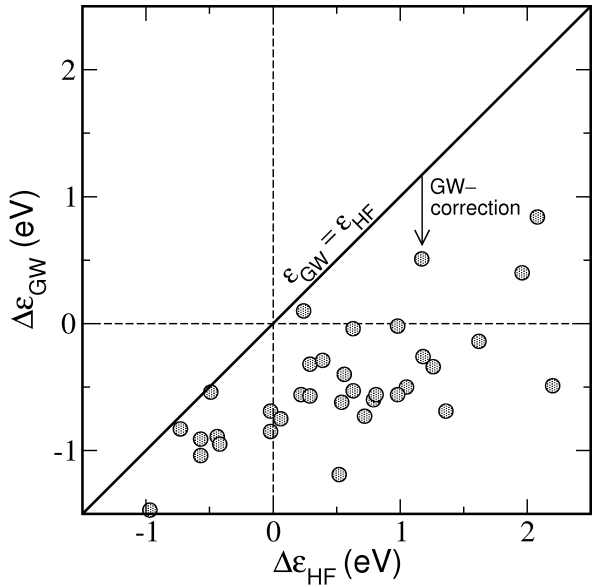}
\caption[cap.wavefct]{\label{fig.hf_vs_gw} The deviation of the
  calculated HOMO energy from the experimental ionization potential in
  GW and HF, respectively. The vertical displacement of points from
  the line $x=y$ gives the difference between the GW and HF energies
  and represents the effect of screening. Notice that the GW
  correction is always negative (corresponding to higher HOMO energy) and that it generally overcorrects the HF energies.
  Also notice that the GW correction is larger for molecules where HF
  presents the largest overestimation of the ionization potential.}
\end{figure}

As can be seen from Fig. \ref{fig.ip}, GW performs better than
Hartree-Fock for the HOMO energy yielding a mean absolute error with
respect to experiments of 0.5 eV compared to 0.81 eV with Hartree-Fock. As expected the difference
between HF and GW is not large on an absolute scale (around 1 eV
on average, see Table \ref{tab2.num}) illustrating the fact that screening is weak in
small molecules. On a relative scale selfconsistent GW improves the
agreement with experiments by almost 30\% as compared to HF.

To gain more insight into the influence of screening on the orbital
energies, we compare in Fig. \ref{fig.hf_vs_gw} the deviation of the
HF and GW energies from $\text{IP}_{\text{exp}}$. The GW self-energy can be
split into the bare exchange potential and an energy-dependent
correlation part
\begin{equation}
  \Sigma_{\text{GW}}(\br, \br';
\varepsilon)=v_x(\br,\br')+\Sigma_{\text{corr}}(\br,\br'; \varepsilon)
\end{equation}
Accordingly the quasiparticle energy can be written as the bare HF
energy and a correction due to the energy-dependent part of the GW
self-energy (the dynamical screening term)
\begin{equation}
  \varepsilon_n^{\text{QP}} = \varepsilon_n^{\text{HF}} + \Delta_n^{\text{GW}}.
\end{equation}
In Fig. \ref{fig.hf_vs_gw} the line $y=x$ corresponds to
$\Delta_n^{\text{GW}}=0$, and the vertical displacement from the line
thus represents the effect of screening on the calculated HOMO energy.
We first notice that the effect of screening is to shift the HOMO
level upwards in energy, i.e. to reduce the ionization potential. This
can be understood by recalling that the Hartree-Fock eigenvalue
represents the energy cost of removing an electron from the HOMO when
orbital relaxations in the final state are neglected (Koopmans'
theorem\cite{grosso}). In Ref. \onlinecite{paperI} we showed, on the
basis of GW and exact calculations for semi-empirical models of
conjugated molecules, that $\Delta_n^{\text{GW}}$ mainly describes the
orbital relaxations in the final state and to a lesser extent accounts
for the correlation energy of the initial and final states. This
explains the negative sign of $\Delta_n^{\text{GW}}$ because the
inclusion of orbital relaxation in the final state lowers the energy
cost of removing an electron. We note that this is different from the
situation in extended, periodic systems where orbital relaxations
vanish and the main effect of the GW self-energy is to account for
correlations in the initial and final states.

\begin{table*}[tp]
  \caption{Experimental ionization potential (first column) and HOMO energy
    calculated using different approximations for exchange and correlation.
     ``X-eig'' refers to a single-particle eigenvalue while ``X-tot'' refers to a total energy difference, $E(N)-E(N-1)$. The G$_0$W$_0$(PBE) energies have been obtained from
    the QP equation while the GW and G$_0$W$_0$ energies are obtained from the DOS in Eq. (\ref{eq.dos}). Last row shows the mean absolute
    error (MAE) with respect to experiments. All energies are in eV.}\label{tab.num}
\begin{tabular}{|l|| *{9}{r@{.}l|}}
\hline
  Molecule       & \multicolumn{2}{c}{Expt.$^{(a)}$} & \multicolumn{2}{|c}{PBE-eig} &
\multicolumn{2}{|c}{PBE0-eig} & \multicolumn{2}{|c}{HF-eig} & \multicolumn{2}{|c}{GW} &
\multicolumn{2}{|c|}{G$_0$W$_0$(HF)} & \multicolumn{2}{|c|}{G$_0$W$_0$(PBE)-QP}& \multicolumn{2}{|c}{MP2$^{(a)}$} & \multicolumn{2}{|c|}{PBE-tot}\\
  \hline \hline
  LiH         &\hspace{6mm}7&90\hspace{4mm} & \hspace{6mm}4&34\hspace{4mm} &
\hspace{6mm}5&81\hspace{4mm} & \hspace{6mm}8&14\hspace{4mm} &
\hspace{6mm}8&0$^{(b)}$\hspace{1mm} & \hspace{6mm}8&2$^{(b)}$\hspace{1mm} &
\hspace{6mm}8&0\hspace{1mm} &
\hspace{6mm}8&20\hspace{1mm} &
\hspace{6mm}8&02 \hspace{1mm} \\
  Li$_2$        & 5&11 & 2&96 & 3&62 & 4&62 & 4&6 & 4&7 & 4&4 & 4&91 & 5&09 \\
  LiF           &11&30 & 6&00 & 8&62 &13&26 &11&7 & 11&2 & 12&0 & 12&64 & 11&87  \\
  Na$_2$        & 4&89 & 2&81 & 3&38 & 4&16 & 4&1 &  4&3 & 4&7 & 4&48 & 4&97\\
  NaCl          & 9&80 & 5&24 & 6&92 & 9&78 & 9&0 &  9&2 & 8&8 & 9&63 & 9&37 \\
  CO            &14&01 & 9&05 &10&98 &14&80 &13&4 & 14&1 & 13&9 & 15&08 & 13&88 \\
  CO$_2$        &13&78 & 9&08 &11&09 &14&50 &13&1 & 13&3 & 13&6 & 14&71 & 13&64 \\
  CS            &11&33 & 7&40 & 9&09 &12&31 &10&8 & 11&7 & 11&0 & 12&58 & 11&31 \\
  C$_2$H$_2$    &11&49 & 7&20 & 8&64 &11&05 &10&6 & 11&1 & 11&2 & 11&04 & 11&39 \\ 
  C$_2$H$_4$    &10&68 & 6&79 & 8&11 &10&11 & 9&8 & 10&4 & 9&6 & 10&18 & 10&67\\ 
  CH$_4$        &13&60 & 9&43 &11&29 &14&77 &14&1 & 14&4 & 14&4$^{(c)}$ & 14&82& 14&10 \\
  CH$_3$Cl      &11&29 & 7&08 & 8&80 &11&68 &11&0 & 11&4 & 11&1 & 11&90 & 11&10\\
  CH$_3$OH      &10&96 & 6&31 & 8&49 &12&14 &10&7 & 10&8 & 10&5 & 12&16 & 10&72 \\
  CH$_3$SH      & 9&44 & 5&60 & 7&09 & 9&50 & 8&8 &  9&0 & 8&4 & 9&73 & 9&29 \\
  Cl$_2$        &11&49 & 7&32 & 9&02 &12&03 &10&9 & 11&3 & 11&5& 12&37 & 11&22 \\
  ClF           &12&77 & 7&90 & 9&88 &13&33 &12&4 & 12&4 & 13&0& 13&63 & 12&48\\  
  F$_2$         &15&70 & 9&43 &12&42 &17&90 &15&2 & 15&2 & 16&2& 18&20 & 15&39 \\   
  HOCl          &11&12 & 6&68 & 8&66 &11&93 &10&6 & 10&8 & 11&0& 12&23 & 10&95 \\
  HCl           &12&74 & 8&02 & 9&78 &12&96 &12&2 & 12&5 & 12&5& 13&02 & 12&71 \\
  H$_2$O$_2$    &11&70 & 6&38 & 8&78 &13&06 &11&0 & 11&1 & 11&1 & 13&00 & 11&18  \\
  H$_2$CO       &10&88 & 6&28 & 8&37 &11&93 &10&4 & 10&5 & 10&6 & 11&97 & 10&80 \\
  HCN           &13&61 & 9&05 &10&67 &13&19 &12&7 & 13&2 & 12&4 & 13&33 & 13&67\\
  HF            &16&12 & 9&61 &12&47 &17&74 &16&0 & 15&6 & 15&7 & 17&35 & 16&27\\
  H$_2$O        &12&62 & 7&24 & 9&59 &13&88 &12&3 & 12&1 & 11&9$^{(d)}$& 13&62 & 12&88 \\
  NH$_3$        &10&82 & 6&16 & 8&11 &11&80 &10&8 & 11&0 & 10&6 & 11&57 & 11&02 \\
  N$_2$         &15&58 &10&28 &12&51 &16&21 &15&1 & 15&7 & 15&6 & 16&41 & 15&39\\
  N$_2$H$_4$    & 8&98 & 5&75 & 7&67 &11&06 & 9&8 & 10&1 & 9&5 & 11&07 & 9&90\\
  SH$_2$        &10&50 & 6&29 & 7&79 &10&48 & 9&8 & 10&1 & 9&9 & 10&48 & 10&38\\
  SO$_2$        &12&50 & 8&08 & 9&96 &13&02 &11&3 & 11&7 & 11&7 & 13&46 & 12&12\\
  PH$_3$        &10&95 & 6&79 & 8&17 &10&38 & 9&9 & 10&3 & 10&0 & 10&50 & 10&39\\
  P$_2$         &10&62 & 7&09 & 8&21 & 9&65 & 9&2 &  9&8 & 9&0 & 10&09 & 10&37\\ 
  SiH$_4$       &12&30 & 8&50 &10&13 &12&93 &12&3 & 12&6 & 12&4$^{(e)}$& 13&25 & 11&95\\
  Si$_2$H$_6$   &10&53 & 7&27 & 8&54 &10&82 &10&2 & 10&6 & 9&9& 11&03 & 10&36\\
  SiO           &11&49 & 7&46 & 9&14 &11&78 &10&9 & 11&2 & 11&3& 11&82 & 11&27\\
  \hline \hline
  MAE & \multicolumn{2}{c|}{-} & 4&35 & 2&55 & 0&81 & 0&5 & 0&4 & 0&5 & 0&82 & 0&24 \\
  \hline 
\end{tabular}
\flushleft
$^{(a)}$From Ref. \onlinecite{nist}. The MP2 calculations use a Gaussian 6-311G$^{**}$ basis set.\\
$^{(b)}$To be compared with the GW value 7.85 and the G$_0$W$_0$(HF) value 8.19 reported in Ref. \onlinecite{stan_eulett}.\\
$^{(c)}$To be compared with the G$_0$W$_0$(LDA) value 14.3 reported in Ref. \onlinecite{grossman01}.\\
$^{(d)}$To be compared with the G$_0$W$_0$(LDA) value 11.94 reported in Ref. \onlinecite{hahn}.\\
$^{(e)}$To be compared with the G$_0$W$_0$(LDA) values 12.7 and 12.66 reported in Refs. \onlinecite{grossman01} and \onlinecite{hahn}, respectively.
\end{table*}

\begin{table*}[tp]
  \caption{Mean absolute deviation between the IPs of the 33 molecules calculated with the different methods and experiment. The mean absolute deviation with respect to experiment coincide with the last row in Table \ref{tab.num}}\label{tab2.num} 
\begin{tabular}{|l|| *{8}{r@{.}l|}}
\hline
  Method       & \multicolumn{2}{c}{Expt.$^{\text{(a)}}$} & \multicolumn{2}{|c}{PBE-eig} &
\multicolumn{2}{|c}{PBE0-eig} & \multicolumn{2}{|c}{HF-eig} & \multicolumn{2}{|c}{GW} & \multicolumn{2}{|c}{G$_0$W$_0$[HF]} & 
\multicolumn{2}{|c|}{MP2$^{\text{(a)}}$} & \multicolumn{2}{|c|}{PBE-tot}\\
  \hline \hline
  Expt.         &\hspace{6mm}0&00\hspace{4mm} & \hspace{6mm}4&35\hspace{4mm} &
\hspace{6mm}2&55\hspace{4mm} & \hspace{6mm}0&81\hspace{4mm} &
\hspace{6mm}0&5\hspace{1mm} & \hspace{6mm}0&4 \hspace{1mm} & \hspace{6mm}0&82 \hspace{1mm} &
\hspace{6mm}0&24\hspace{1mm}\\
  PBE        & 4&35 & 0&00 & 1&79 & 4&90 & 3&9 & 4&1 & 4&99 & 4&27\\
  PBE0           & 2&55  & 1&79 & 0&00 & 3&11 & 2&1 & 2&3 & 3&20 & 2&48\\
  HF        & 0&81 & 4&90 & 3&11 & 0&00 & 1&0 & 0&8 &  0&17 & 0&80\\
  GW          & 0&5 & 3&9 & 2&1 & 1&0 & 0&00 & 0&3 & 1&1 & 0&4\\
  G$_0$W$_0$[HF] & 0&4 & 4&1 & 2&3 & 0&8 & 0&3 & 0&00 & 0&9 & 0&3 \\
  MP2            & 0&82 & 4&99 & 3&20 & 0&17 & 1&1 & 0&9 & 0&00 & 0&84\\
  PBE-tot        & 0&24 & 4&27 & 2&48 & 0&80 & 0&4 & 0&3 & 0&84 & 0&00\\
  \hline \hline
\end{tabular}
\flushleft
$^{(a)}$Data taken from Ref. \onlinecite{nist}.
\end{table*}

In Table \ref{tab.num} we list the calculated HOMO energy for each of
the 33 molecules. In addition to selfconsistent GW we have performed
one-shot G$_0$W$_0$ calculations based on the HF and PBE Green's
function, respectively. The best agreement with experiment is obtained
for G$_0$W$_0$[HF]. This is because the relatively large Hartree-Fock
HOMO-LUMO gap reduces the (over-)screening described by the resulting GW self-energy.
There are not many GW calculations for molecules available in the
litterature. Below Table \ref{tab.num} we list the few we have found.
As can be seen they all compare quite well with our results given the
differences in the implementation of the GW approximation.

For comparison we have included the HOMO energy predicted by second
order M{\o}ller-Plesset theory (MP2) [taken from Ref.
\onlinecite{nist}] with a Gaussian 6-311G$^{**}$ basis set. These are
generally very close to our calculated HF values, with a tendency to
lower energies which worsens the agreement with experiment slightly as compared to HF.

We have also calculated the DFT-PBE total energy difference between
the neutral and cation species, $E(N)-E(N-1)$, see last column of
Table \ref{tab.num}. This procedure leads to IPs in very good agreement with the experimental values (MAE of 0.24 eV). 
We stress that although this method is superior to the GW method for the IP of the small molecules studied here, 
it can yield only the HOMO and LUMO levels while higher excited states
are inaccessible. Moreover it applies only to isolated systems and cannot
be directly used to probe QP levels of e.g. a molecule on a surface. 

In Table \ref{tab2.num} we provide an overview of the comparative performance of the different methods. Shown is 
the mean average deviation between the IPs
calculated with the different methods as well as the experimental
values. Note that the numbers in the experiment row/column are the
same as those listed in the last row of Table \ref{tab2.num}.

\begin{figure}[!ht]
\includegraphics[width=0.9\linewidth]{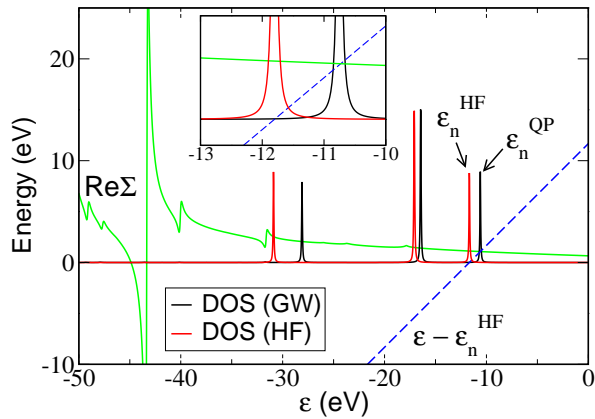}
\caption[cap.wavefct]{\label{fig.nh3} (Color online) Density of states
  for the NH$_3$ molecule calculated in HF and GW, respectively.
  Arrows mark the level corresponding to the HOMO in the two
  calculations. The intersection between the line
  $y=\varepsilon-\varepsilon_n^{\text{HF}}$ and the real part of
  $\langle
  \psi^0_{\text{HOMO}}|\Sigma_{\text{corr}}(\varepsilon)|\psi_{\text{HOMO}}^0\rangle$
  (green curve) determines the position of the GW level.}
\end{figure}

\subsection{Linearized quasiparticle equation}
In the conventional GW method the full Green function of Eq.
(\ref{eq.gr}) is not calculated.  Rather one obtains the quasiparticle
energies from the quasiparticle equation
\begin{equation}\label{eq.qpeq}
  \varepsilon_n^{\text{QP}} = \varepsilon_n^{0} + Z_n \langle
  \psi_n^{0}|\Sigma_{\text{GW}}(\varepsilon_n^{0})-v_{\text{xc}}|\psi_n^{0}
\rangle.
\end{equation}
where $\psi_n^{0}$ and $\varepsilon_n^{0}$ are eigenstates and
eigenvalues of an approximate single-particle Hamiltonian (often the
LDA Hamiltonian), and
\begin{equation}
  Z_n =\Big [1-\frac{\partial \langle
\psi_n^{\text{DFT}}|\Sigma_{\text{GW}}(\varepsilon)|\psi_n^{0}
\rangle}{\partial
\varepsilon}\Big |_{\varepsilon_n^{0}}\Big ]^{-1}.
\end{equation}
Moreover the GW self-energy is evaluated non-selfconsistently from the
single-particle Green function, i.e.  $\Sigma_{\text{GW}}=
iG_0W[G_0]$, with $G_0(z)=(z-H_{0})^{-1}$.

The quasiparticle equation (\ref{eq.qpeq}) relies on the assumption
that off-diagonal matrix elements, $\langle \psi_n^{0} |
\Sigma_{\text{GW}} (\varepsilon_n^{0}) - v_{\text{xc}} |\psi_m^{0}
\rangle$, can be neglected, and that the frequency dependence of
$\Sigma_{\text{GW}}$ can be approximated by its first order Taylor
expansion in a sufficiently large neighborhood of $\varepsilon_n^{0}$.
We have found that these two assumptions are indeed fullfilled for the
molecular systems studied here. More precisely, for the GW and
G$_0$W$_0$(HF) self-energies, the QP energies obtained from
Eq. (\ref{eq.qpeq}) are always very close to the peaks in the density
of states Eq. (\ref{eq.dos}). We emphasize that this result could well
be related to the rather large level spacing of small molecules, and
may not hold for extended systems.  An example
is presented in Fig. \ref{fig.nh3} which shows the full HF and GW
density of states for NH$_3$ together with the real part of $\langle
\psi^{0}_{\text{HOMO}} | \Sigma_{\text{corr}}(\varepsilon) |
\psi_{\text{HOMO}}^{0}\rangle$. As explained in the following section
this is not quite the case for the G$_0$W$_0$(PBE) calculations.

\subsection{$G_0$-dependence}
As stated in the previous section the GW and G$_0$W$_0$(HF) energies
can be obtained either from the full spectral function or from the QP
equation. In this case, returning to Table \ref{tab.num}, we see that
G$_0$W$_0$(HF) yields systematically larger IPs than GW. This is easy
to understand since $G_{\text{HF}}$ describes a larger HOMO-LUMO gap
than $G_{\text{GW}}$, and therefore produces less screening. When the
PBE rather than the HF Green function is used to evaluate the GW
self-energy, we find that the spectral function obtained from
Eq. (\ref{eq.gr}) does not resemble a simple discrete spectrum. In
fact the peaks are significantly broadened by the imaginary part of
$\Sigma_{\text{GW}}$ and it becomes difficult to assign precise values
to the QP energies. Apart from the spectral broadening, the molecular
gap is significantly reduced with respect to its value in the GW and
G$_0$W$_0$(HF) calculations. Both of these effects are due to the very
small HOMO-LUMO gap described by $G_{\text{PBE}}$ which leads to
severe overscreening and QP life-time reductions. A similar effect was
observed by Ku and Eguiluz in their comparison of GW and
G$_0$W$_0$(LDA) for Si and Ge crystals\cite{ku}. 

The problems encountered when attempting to solve the Dyson equation
(\ref{eq.gr}) using the G$_0$W$_0$(PBE) self-energy occur due to the
large mismatch between $\varepsilon_n^{\text{PBE}}$ and
$\varepsilon_n^{\text{QP}}$. On the other hand, in the QP equation,
the GW self-energy is evaluated at $\varepsilon_n^0$ rather than
$\varepsilon_n^{\text{QP}}$. As a consequence the unphysical
broadening and overscreening is avoided and a well defined QP energy
can be obtained (last column in Table \ref{tab.num}).

To summarize, $G_0$ can have a very large effect on the QP spectrum
when the latter is obtained via the Dyson equation (\ref{eq.gr}). In
particular, the use of a $G_0$ with a too narrow energy gap (as e.g.
the $G_{\text{PBE}}$) can lead to unphysical overscreening and
spectral broadening. When the QP levels are obtained from the QP
equation, the $G_0$-dependence is less pronounced because
$\Sigma_{\text{GW}}[G_0]$ is evaluated at $\varepsilon_n^0$ which is
consistent with $G_0$.

The self-consistent GW spectrum is independent of the choice of $G_0$,
but the number of iterations required to reach self-consistency is
less when based on $G_\text{HF}$.

\subsection{Basis set convergence}\label{sec.conv}
\begin{figure}[!ht]
\includegraphics[width=0.75\linewidth]{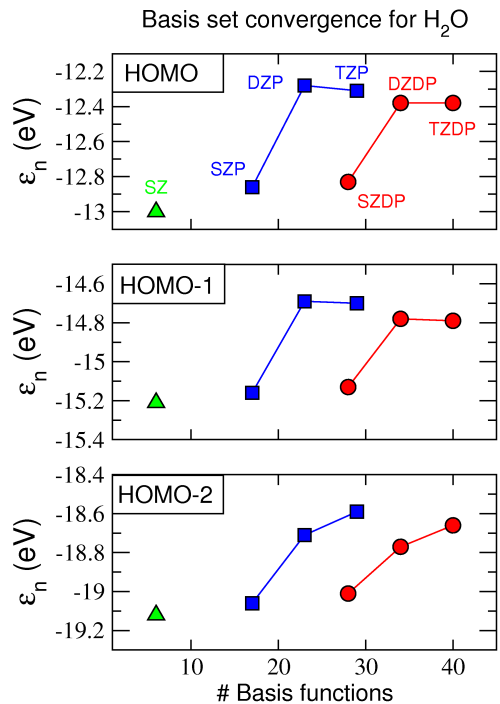}
\caption[cap.wavefct]{\label{fig.convH2O} (Color online) Convergence
  of the three highest occupied levels of H$_2$O obtained from GW
  calculations with different sizes of the augmented Wannier function
  basis. SZ denotes the Wannier function basis, while e.g. DZDP
  denotes the Wannier basis augmented by one extra radial function per
  valence state and two sets of polarization functions.}
\end{figure}

\begin{figure}[!ht]
  \includegraphics[width=0.75\linewidth]{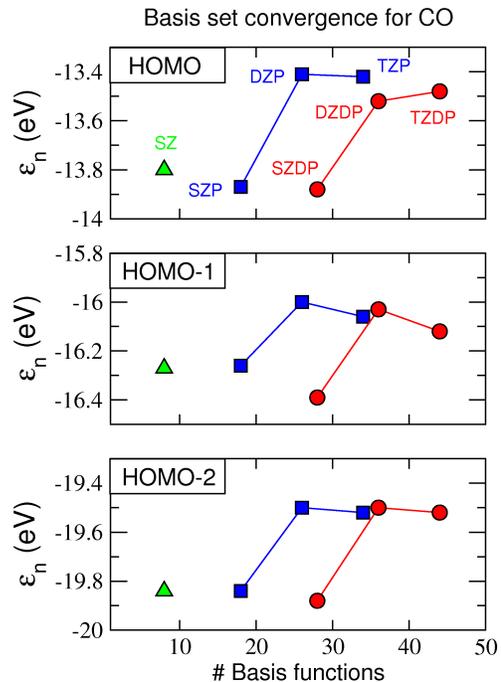}
  \caption[cap.wavefct]{\label{fig.convCO} (Color online) Same as Fig.
    \ref{fig.convH2O} but for CO.}
\end{figure}

In Figs. \ref{fig.convH2O} and \ref{fig.convCO} we show the energy of
the three highest occupied molecular orbitals of H$_2$O and CO
obtained from selfconsistent GW using various sizes of the augmented
Wannier basis. Clearly, the polarization functions have relatively
little influence on the QP energies while the first set of additional
zeta functions reduce the QP energies by up to 0.5 eV. The differences
between DZP and TZDP are less than 0.15 eV for all the levels which
justifies the use of DZP basis.

We have also compared the eigenvalues obtained from selfconsistent HF
calculations using the DZP augmented Wannier basis to accurate HF
calculations performed with the real-space code GPAW\cite{GPAW}. Here
we obtain a MAE of 0.09 eV for the energy of the HOMO level of the 33
molecules.

\section{Conclusions}\label{sec.conclusions}
As the range of systems to which the GW method is being applied
continues to expand it becomes important to establish its performance
for other systems than the solids.  In this work we have discussed
benchmark GW calculations for molecular systems.

The GW calculations were performed using a novel scheme based on the
PAW method and a basis set consisting of Wannier functions augmented
by numerical atomic orbitals. We found that a basis corresponding to
double-zeta with polarization functions was sufficient to obtain GW
energies converged to within ~0.1 eV (compared to triple-zeta with
double polarization functions). The GW self-energy was calculated on
the real frequency axis including its full frequency dependence and
off-diagonal elements. We thereby avoid all of the commonly used
approximations, such as the the plasmon pole approximation, the
linearized quasiparticle equation and analytical continuations from
imaginary to real frequencies, and thus obtain a direct and unbiased
assessment of the GW approximation itself. We found that the inclusion
of valence-core exchange interactions, as facilitated by the PAW
method, is important and affect the HF/GW HOMO levels by -1.2 eV on
average.

The position of the HOMO for a series of 33 molecules was calculated
using fully selfconsistent GW, single-shot G$_0$W$_0$, Hartree-Fock,
DFT-PBE0, and DFT-PBE. Both PBE and PBE0 eigenvalues grossly
overestimate the HOMO energy with a mean absolute error (MAE) with
respect to the experimental ionization potentials (IP) of 4.4 and 2.5
eV, respectively. Hartree-Fock underestimates the HOMO energy but
improves the agreement with experiments yielding a MAE of 0.8 eV. GW
and G$_0$W$_0$ overcorrects the Hartree-Fock levels slightly leading
to a small overestimation of the HOMO energy with a MAE relative to
experiments of 0.4-0.5 eV. This shows that although screening is a
weak effect in molecular systems its inclusion at the GW level
improves the electron removal energies by 30-50\% relative to the unscreened Hartree-Fock. The best IPs were obtained from one-shot G$_0$W$_0$ calculations starting from the HF Green's function where the overscreening is least severe. Very similar conclusions were
reached by comparing GW, G$_0$W$_0$ and HF to exact diagonalization for
conjugated molecules described by the semi-empirical PPP
model.\cite{paperI}

\section{Acknowledgments}
We thank Mikkel Strange for useful discussions and assistance with the
projected Wannier functions.  We acknowledge support from the Danish
Center for Scientific Computing and The Lundbeck Foundation's Center
for Atomic-scale Materials Design (CAMD).

\appendix
\section{The GW self-energy}
\label{sec:GWselfenergy}

Let $U$ denote the rotation matrix that diagonalizes the pair orbital
overlap $S_{ij,kl} = \braket{n_{ij}}{n_{kl}}$, i.e. $U^\dagger S U =
\sigma I$. The columns of $U$ are truncated to those which have
corresponding eigenvalues $\sigma_q < 10^{-5} a_0^{-3}$. We then only
calculate the reduced number of Coulomb elements
\begin{equation}
  V_{qq'} = \bra{n_q} \frac{1}{\rr} \ket{n_{q'}},
\end{equation}
where $n_q(\br)$ are the optimized pair orbitals
\begin{equation}
  n_q(\br) = \sum_{ij} n_{ij}(\br) U_{ij,q} / \sqrt{\sigma_q},
\end{equation}
which are mutually orthonormal, i.e. $\braket{n_q}{n_{q'}} =
\delta_{qq'}$.

Determining the GW self-energy proceeds by calculating first the full
polarization matrix in the time domain
\begin{align}
  P^<_{ij,kl}(t) &= 2 i G^<_{ik}(t) G^{>*}_{jl}(t),\label{eq:P-lesser}\\ 
 P^>_{ij,kl}(t) &= P^{<*}_{ji,lk}(t).
\end{align}
The factor 2 appears for spin-paired systems from summing over spin indices.
This is then downfolded to the reduced representation
\begin{equation}
  P^{\lessgtr}_{qq'} = \sum_{ij,kl} \sqrt{\sigma_q} U^*_{ij,q} P^{\lessgtr}_{ij,kl}
U_{kl,q'} \sqrt{\sigma_{q'}}.
\end{equation}
The screened interaction can be determined from the lesser and greater
polarization matrices, and the static interaction $V_{qq'}$, via the
relations
\begin{align}
  P^r(t) &= \theta(t) \left( P^>(t) - P^<(t) \right),\\
  W^r(\varepsilon) &= [I - V P^r(\varepsilon)]^{-1}V,\\
  W^>(\varepsilon) &= W^r(\varepsilon) P^>(\varepsilon) W^{r\dagger}(\varepsilon),\\
  W^<(\varepsilon) &= W^>(\varepsilon) - W^r(\varepsilon) + W^{r\dagger}(\varepsilon),
\end{align}
where all quantities are matrices in the optimized pair orbital basis
and matrix multiplication is implied. We obtain the screened
interaction in the original orbital basis from
\begin{equation}
  W^{\lessgtr}_{ij,kl}(t) \approx \sum_{qq'} U_{ij,q} \sqrt{\sigma_q}
W^{\lessgtr}_{qq'}(t) \sqrt{\sigma_{q'}} U_{kl,q'}^*,
\end{equation}
which is an approximation due to the truncation of the columns of U.
Finally the GW self-energy can be determined by
\begin{align}
  \Sigma^{\lessgtr}_{\text{GW,}ij}(t) &= i \sum_{kl} G^{\lessgtr}_{kl}(t)
W^{\lessgtr}_{ik,jl}(t)\\
  \Sigma^r_\text{GW}(t) &= \theta(t) \left( \Sigma^>_\text{GW}(t) -
\Sigma^<_\text{GW}(t) \right)+\delta(t)\Sigma_x.
\end{align}
The exchange and Hartree potentials are given by
\begin{align}
  \Sigma_{x,ij} &= i\sum_{kl}V_{ik,jl}G^<_{kl}(t=0)\\
  \Sigma_{\text{H},ij} &= -2i\sum_{kl}V_{ij,kl}G^<_{kl}(t=0) 
\end{align}

The Green functions are given by 
\begin{align}
  G^r(\varepsilon)=&[(\varepsilon+i\eta)S-H_{\text{KS}}+ v_{\text{xc}} -\Delta
v_{\text{H}}-\Sigma_{\text{GW}}^r(\varepsilon)]^{-1}\\
  G^<(\varepsilon)=&-f_{\text{FD}}(\varepsilon-\mu)[G^r(\varepsilon)-G^r(\varepsilon)^\dagger]\\
  G^>(\varepsilon)=&(1-f_{\text{FD}}(\varepsilon-\mu))[G^r(\varepsilon)-G^r(\varepsilon)^\dagger]\label{eq:g-greater}
\end{align}
where $f_{\text{FD}}(\varepsilon-\mu)$ is the Fermi-Dirac function and
$\Delta v_{\text{H}} =
\Sigma_{\text{H}}[G]-\Sigma_{\text{H}}[G_{\text{DFT}}]$ is the
difference between the GW Hartree potential and the DFT Hartree
potential. For self-consistent calculations, equation
\eqref{eq:P-lesser}-\eqref{eq:g-greater} are iterated untill
convergence in $G$.


\bibliographystyle{apsrev}

\begin{thebibliography}{100}
\bibitem{jarnak_comment} According to Janak's theorem\cite{jarnak},
  for finite systems the position of the HOMO in exact Kohn-Sham
  theory should coincide with the ionization potential. However, for
  practical exchange-correlation functionals this is far from being the
  case\cite{perdew_zunger,pemmaraju07}.
\end{thebibliography}

\end{document}